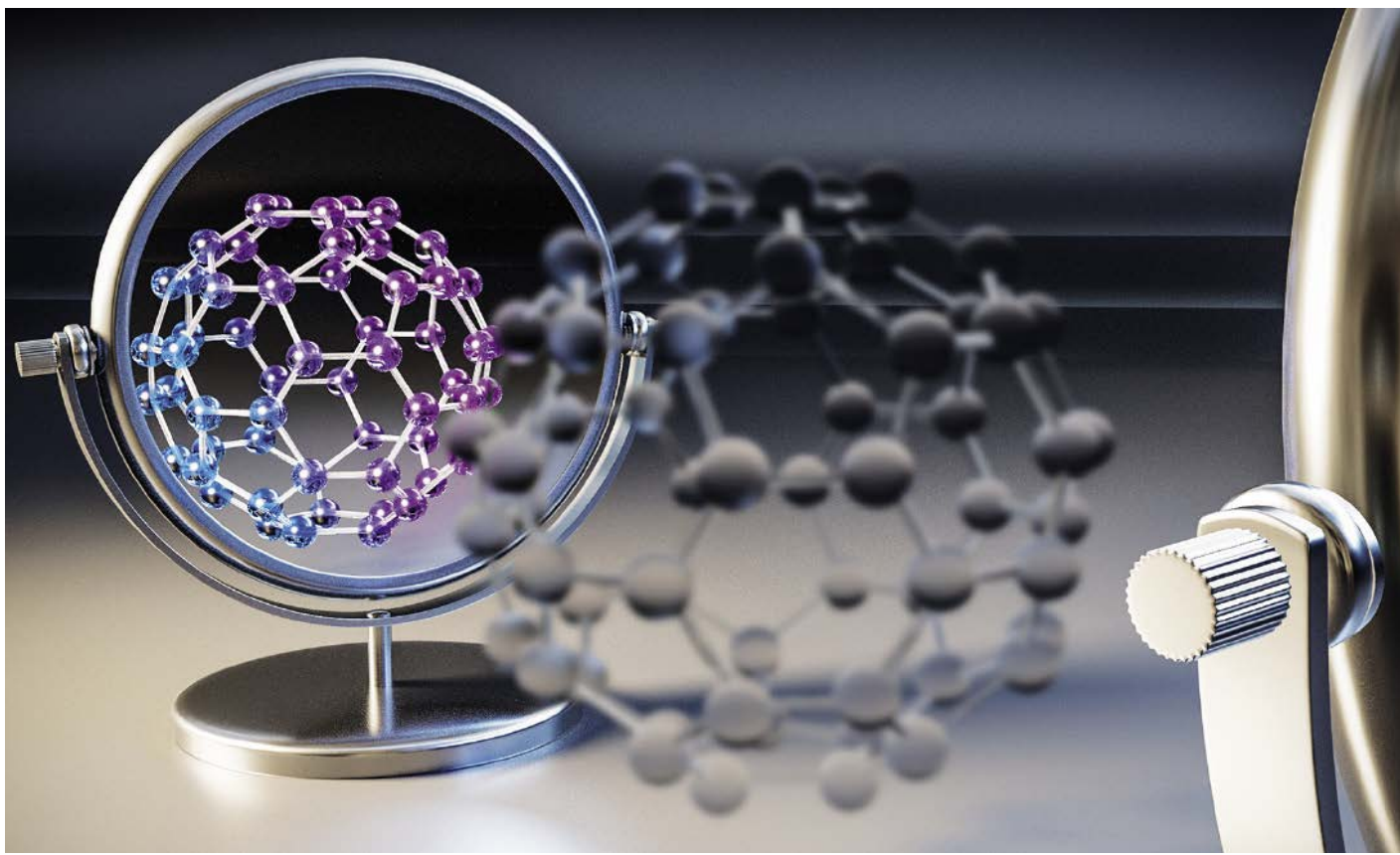

# Seeing the Unseen

## Boosted Absorption Imaging and Spectroscopy Using a Scanning Microresonator

*Jonathan Noé[1,2], Michael Förg[1,2], Manuel Nutz[1,2], Florian Steiner[2], Rute Fernandes[2], Ines Amersdorffer[2], David Hunger[3,4], Thomas Hümmer[1,2]*

**O**ptical detection of nanoscale objects without relying on fluorescence is a current challenge due to their extremely weak interaction with light. Resonator-enhanced absorption microscopy is a novel tool to heavily boost the light-matter interaction. It combines highly sensitive, spatially resolved imaging with spectroscopy, offering a competitive alternative to established methods for the detection and analysis of nanoscale systems in nanotechnology, material design, and the life sciences.

### Current Limitations of Microscopy for Nanomaterials

Characterizing nanoscale objects is pivotal in the fields of nanotechnology and the life sciences. Currently, the most used methods to detect and characterize the optical properties of nanoscale materials are fluorescence and Raman scattering. Raman scattering, however, requires very high laser intensities to characterize nanoscale matter, which often damages or destroys the sample. Fluorescence measurements, on the other hand, are limited to only a small fraction of all nanoscale materials that show intrinsic fluorescence. Alternatively, detection by labeling with fluorescent dyes is restricted to nanoscale objects that are viable for labeling and does not provide information about the inherent optical properties of the sample. Furthermore, dyes are subject to photobleaching and therefore limit the time for investigation.

Since light absorption is an inherent property of all materials, absorption microscopy is a promising alternative for the characterization of nanomaterials. However, measuring the absorption signal of individual nanoobjects is challenging, since the absorption cross-section scales linearly with the volume of the object. Only a handful of imaging techniques have demonstrated the ability to image such weak absorption signals and only on a laboratory scale, where the required sensitivity is achieved by various elaborate noise rejection techniques [1-3]. As an alternative to these approaches, straightforward highly sensitive absorption measurements can be accomplished by signal enhancement via optical microresonators.

### Resonator-Enhanced Absorption Microscopy

Optical resonators consist of two opposing highly reflective mirrors, between which light forms a standing wave. In other words, light travels back and forth up to a million times before exiting the resonator. As a result, the interaction between light and matter within this resonator can be amplified by many orders of magnitude. Pioneered in quantum optics this technology is known for its ultra-sensitive sensing capability since the many roundtrips of the light in the resonator enhance various photophysical processes (e.g., absorption, fluorescence, scattering, dispersion) [4]. Nonetheless, the large resonator mode of conventional Fabry-Perot cavities and the lack of control over its position relative to the sample, make them unsuitable for imaging.

This limitation was overcome by the invention of microscopic mirrors [5]. These atomically smooth concave micromirrors are typically fabricated by laser ablation on the end facet of an optical fiber which is subsequently coated with a highly reflective dielectric coating. Combined with a planar macro-scopic sample mirror at a distance of only a few micrometers, a resonator is formed. This yields an almost diffraction-limited mode waist on the surface of the planar mirror (Fig. 1) and thereby offers the spatial resolution required for imaging [6,7]. Scanning the micromirror over the macroscopic mirror, an image of the sample can be obtained, where light has interacted with the sample several thousands of times within each pixel [7]. Due to the many round trips of the light in the resonator, even weakly absorbing nanoscale objects on the sample mirror lead to an easily detectable reduction in the resonator transmission, making even minuscule absorption (0.0001%) visible.

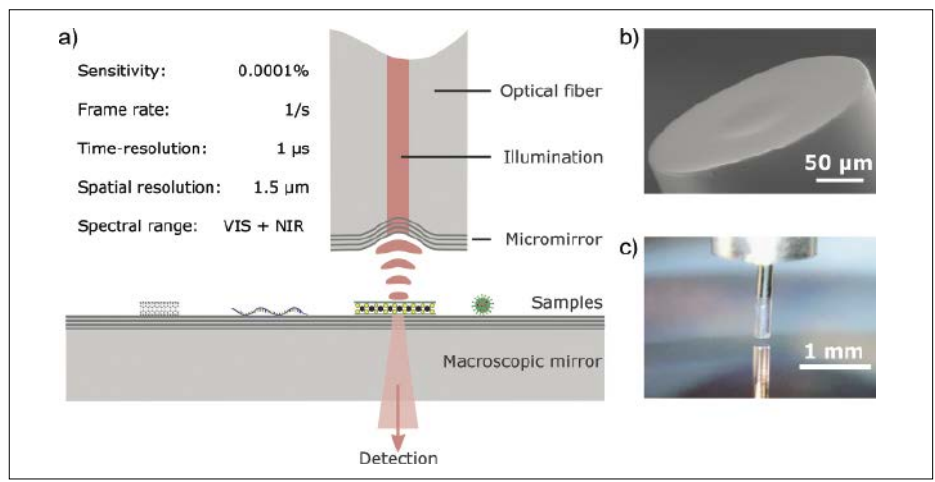

*Fig. 1: (a) Specifications and schematic drawing of the resonator-enhanced absorption microscope formed by a concave micromirror on the end facet of an optical fiber (top) and a macroscopic sample mirror (bottom). Weakly absorbing samples are deposited on top of the macroscopic mirror. (b) Scanning electron microscopy image of the fiber end facet after laser treatment. (c) Photograph of the micromirror and its reflection in the macroscopic mirror while forming a resonator.*

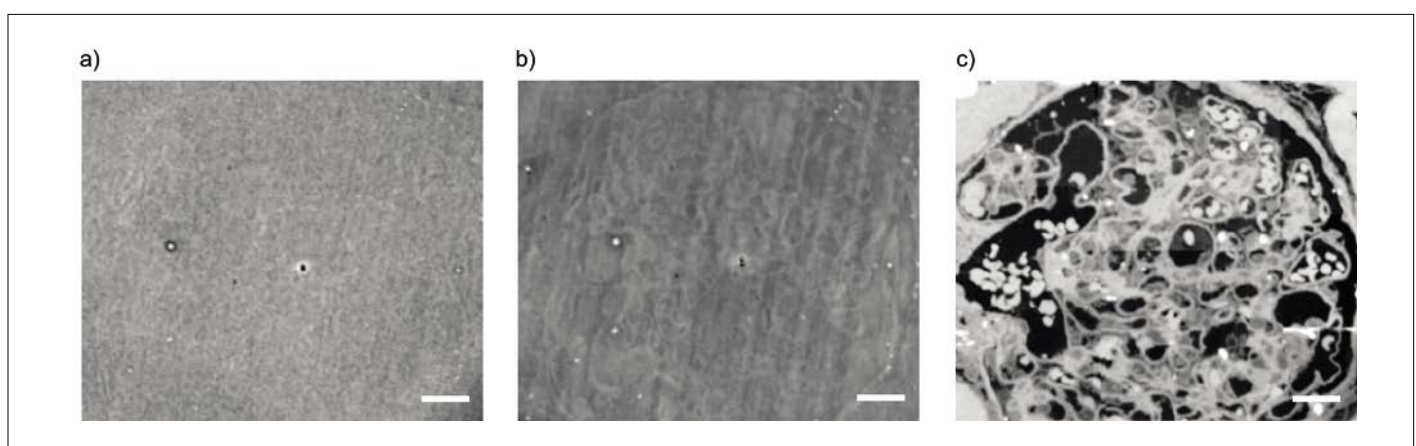

*Fig. 2: Image of an unstained ultra-thin (~ 80 nm) section of a human kidney biopsy in (a) a brightfield, (b) a phase contrast light microscope, and (c) a resonator microscope. Scale bars, 20 µm. (Samples provided by Prof. Dr. med. Stefan Porubsky, Institute for Pathology, University Medical Center of the Johannes Gutenberg University, Mainz)*

The power of this new technique is illustrated in Figure 2 showing the comparison between images of an unstained ultra-thin (~ 80 nm) section from a human kidney biopsy acquired by brightfield (Fig. 2a), phase contrast (Fig. 2b), and resonator microscopy (Fig. 2c). The strong light-matter interaction achieved with the microresonator leads to a significantly increased contrast, making weakly absorbing biological structures visible. This level of contrast is usually only achieved in a conventional light microscope by imaging a semi-thin (100-times thicker) section that has additionally been treated with a contrast agent.

## Highly Sensitive Imaging and Spectroscopy in the Resonator Microscope

The potential and versatility of resonator microscopy have been demonstrated on various nanoscale objects, ranging from the ultrathin sections in Figure 2 over macromolecules and atomically thin semiconductors down to atomistic defects. Carbon nanotubes (Fig. 3a) for example are an important building block of future nanoscale optoelectronic devices. Their absorption in the near-infrared could be imaged for the first time on a single tube level by resonator-enhanced microscopy [6].

By tuning the illumination wavelength inside the resonator microscope, it is possible to measure the full absorption spectrum of a weakly absorbing material. Since only light meeting the resonance condition can enter the resonator, the microscope acts as a very sharp spectral filter.

Figure 3b shows an example of the absorption spectra of a transition metal dichalcogenide monolayer, where parts were treated with different doses of an ion beam to create atomic defects in the structure (marked areas) [8]. Even very low defect concentrations could be imaged (Fig. 3b, inset) and spectrally characterized with the resonator microscope.

Not only the wavelength of the illumination light but also its polarization can be freely chosen. An example of the use of ultrasensitive imaging combined with polarization control is shown in Figure 3c [9]. Polarization-dependent non-fluorescent excitations in a two-dimensional semiconductor heterobilayer, which are connected to crystal strain in the structure, are resolved spectroscopically and by imaging (Fig. 3c, inset).

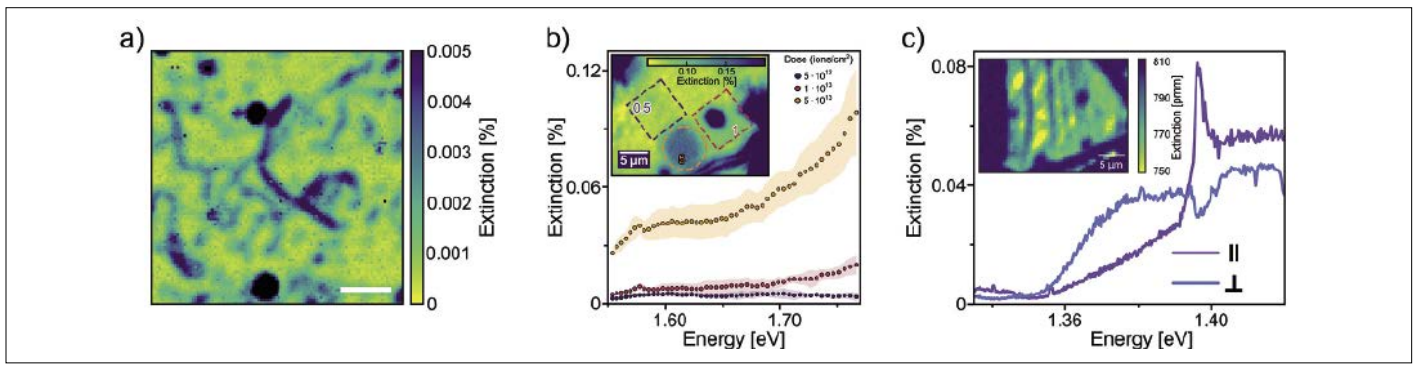

*Fig. 3: (a) Extinction map of single carbon nanotubes dispersed directly on top of the macroscopic mirror of the resonator, scale bar: 20 μm. (b) Extinction spectra of different defect densities in a two-dimensional semiconductor. Inset: Extinction map showing areas of different defect densities. (c) Extinction spectra of a two-dimensional semiconductor heterostructure at two orthogonal illumination polarizations. Inset: Extinction map for one polarization direction.*

## Conclusion

Smaller and smaller building blocks in nanotechnology, material design, and the life sciences create the demand for better detection and characterization of nanoscale matter. Resonator-enhanced imaging and spectroscopy allow to access one of the most fundamental properties of matter, namely absorption, even for extremely weakly absorbing materials. With resulting sensitivity orders of magnitude below the shot noise limit of conventional absorption microscopes, many nanoscopic objects can be optically detected for the first time. The potential of this technique was already demonstrated by the visualization and characterization of macromolecules, atomically thin semiconductors, and thin films utilizing imaging, spectroscopy, and polarization contrast.

## Acknowledgments

This work was supported by the Bundesministerium für Wirtschaft und Klimaschutz (BMWK, Federal Ministry for Economic Affairs and Climate Action), and Europäischer Sozialfonds (ESF, European Social Fund) within an EXIST Transfer of Research (Qlibri, Grant No. 03EFPBY231), the European Commission within the program Horizon 2020 "Fiber Nano-Engineering" (FINE, Grant No. 101034604), and the Bundesministerium für Bildung und Forschung (BMBF, Federal Ministry of education and research) within a GO-BIO Initial project (HAL2, Grant No. 16LW0124).

*Affiliations*
[1]Qlibri GmbH, Munich, Germany
[2]Fakultät für Physik, Center for Nanoscience (CeNS), Munich Center for Quantum Science and Technology (MCQST), Ludwig-Maximilians-Universität, Munich, Germany
[3]Physikalisches Institut, Karlsruhe Institute of Technology, Karlsruhe, Germany
[4]Institut für Quantenmaterialien und Technologien, Eggenstein-Leopoldshafen, Germany

**Contact**
Dr. Jonathan Noé (CEO)
Qlibri GmbH
Munich, Germany
noe@qlibri.eu

[1] References: https://bit.ly/IM-Noe

## References

1. Boyer, D., Tamarat, P., Maali, A., Lounis, B. & Orrit, M. Photothermal Imaging of Nanometer-Sized Metal Particles Among Scatterers. *Science* **297**, 1160-1163 (2002). https://doi.org:doi:10.1126/science.1073765
2. Arbouet, A., Christofilos, D., Del Fatti, N., Vallée, F., Huntzinger, J. R., Arnaud, L., Billaud, P. & Broyer, M. Direct Measurement of the Single-Metal-Cluster Optical Absorption. *Physical Review Letters* **93** (2004). https://doi.org:10.1103/physrevlett.93.127401
3. Ortega-Arroyo, J. & Kukura, P. Interferometric scattering microscopy (iSCAT): new frontiers in ultrafast and ultrasensitive optical microscopy. *Physical Chemistry Chemical Physics* **14**, 15625 (2012). https://doi.org:10.1039/c2cp41013c
4. G. Gagliardi & H.-P. Loock eds *Cavity-Enhanced Spectroscopy and Sensing*. 1 edn, (Springer 2014).
5. Hunger, D., Steinmetz, T., Colombe, Y., Deutsch, C., Hänsch, T. W. & Reichel, J. A fiber Fabry–Perot cavity with high finesse. *New Journal of Physics* **12**, 065038 (2010). https://doi.org:10.1088/1367-2630/12/6/065038
6. Hümmer, T., Noe, J., Hofmann, M. S., Hänsch, T. W., Högele, A. & Hunger, D. Cavity-enhanced Raman microscopy of individual carbon nanotubes. *Nature Communications* **7**, 12155 (2016). https://doi.org:10.1038/ncomms12155
7. Mader, M., Reichel, J., Hänsch, T. W. & Hunger, D. A scanning cavity microscope. *Nature Communications* **6**, 7249 (2015). https://doi.org:10.1038/ncomms8249
8. Sigger, F., Amersdorffer, I., Hötger, A., Nutz, M., Taniguchi, T., Watanabe, K., Förg, M., Noe, J., Finley, J. J., Högele, A., Hänsch, T. W., Holleitner, A. W., Hunger, D., Hümmer, T. & Kastl, C. Ultra-sensitive extinction measurements of optically active defects in monolayer MoS2. *The Journal of Physical Chemistry Letters* (2022). https://doi.org:https://doi.org/10.1021/acs.jpclett.2c02386
9. Förg, M. *Confocal and cavity-enhanced spectroscopy of semiconductor van der Waals heterostructures* Doctoral thesis, Ludwig-Maximilians-Universität München, (2020).